\documentstyle[prb,aps,epsf]{revtex}
\begin{document}
% for two column  activate the line below...                
\twocolumn[\hsize\textwidth\columnwidth\hsize\csname@twocolumnfalse\endcsname
\title{Theory of Interplay of Nuclear Magnetism and Superconductivity in AuIn$_2$}
\author{M. L. Kuli\'c$^{1,2}$, A.I. Buzdin$^2$, and L.N. Bulaevskii $^3$}
\address{$^{1}$Max-Planck-Institut f\"{u}r Festk\"{o}rperforschung, Heisenbergstr. 1,
70569 Stuttgart, Germany\\
$^{2}$Centre de Physique Th\'{e}orique et de Mod\'{e}lisation, Universit\'{e}
Bordeaux I, CNRS-URA 1537 Gradignan Cedex, France\\
$^{3}$Los Alamos National Laboratory, Los Alamos, NM 87545 USA}
\date{\today}
\maketitle

\begin{abstract}
The recently reported \cite{Pobell3} coexistence of a magnetic order, with
the critical temperature $T_M=35$ $\mu$K, and superconductivity, with the
critical temperature $T_S=207$ $m$K, in AuIn$_2$ is studied theoretically.
It is shown that superconducting (S) electrons and localized nuclear
magnetic moments (LM's) interact dominantly via the contact hyperfine (EX)
interaction, giving rise to a spiral (or domain-like) magnetic order in
superconducting phase. The electromagnetic interaction between LM's and S
electrons is small compared to the EX one giving minor contribution to the
formation of the oscillatory magnetic order. In clean samples ($l>\xi _0$)
of AuIn$_2$ the oscillatory magnetic order should produce a line of nodes in
the quasiparticle spectrum of S electrons giving rise to the power law
behavior. The critical field $H_c(T=0) $ in the coexistence phase is reduced
by factor two with respect to its bare value. \\
{\bf e-mail}: kulic@audrey.mpi-stuttgart.mpg.de
\end{abstract}

% for two column  activate the line below... 
]
\narrowtext

The problem of the coexistence of magnetic (M) order and superconductivity
(S) is a long-standing one, which was first considered in 1956 theoretically
by V. L. Ginzburg\cite{Ginzburg}, and then intensively discussed after the
discovery of the ternary rare earth (RE) compounds (RE)Rh$_{4}$B$_{4}$ and
(RE)Mo$_{6}$X$_{8}$ (X=S,Se) \cite{BuBuKuPa1,Maple}. In many of these
compounds both ferromagnetic (F) and antiferromagnetic (AF) orderings, which
coexist with S, have been observed \cite{BuBuKuPa1,Maple}. It turns out that
S and AF orderings coexist in several of these compounds \cite{Buzdin}
usually down to $T=0$ K, while S and modified F (spiral) orderings coexist
only in limited temperature interval in ErRh$_{4}$B$_{4}$, HoMo$_{6}$S$_{8}$
and HoMo$_{6}$Se$_{8}$, due to their antagonistic characters \cite{BuBuKuPa1}%
. A general theory of magnetic superconductors has been developed in Refs. 
\onlinecite
{BuBuKuPa1,BuBuKuPa2,BuRuKu}, where possibilities for the coexistence of S
and spiral or domain-like magnetic order (which is the modified F order in
the presence of superconductivity) have been elaborated quantitatively by
including the exchange and electromagnetic interaction of superconducting
electrons and localized magnetic moments (LM's). To the similar conclusion
came also Blount and Varma \cite{BloVar} by taking into account the
electromagnetic interaction only. Note, that some heavy fermions UPt$_{3}$,
URu$_{2}$Si$_{2}$ etc. show the coexistence of the AF and S orderings, while
S and oscillatory M order coexist in quaternary intermetallic compounds
(RE)Ni$_{2}$B$_{2}$C, see Ref.~\onlinecite{Chang}.

However, until recently it was impossible to investigate the interplay of S
and nuclear magnetic order, because of lack of suitable materials. Thanks to
the pioneering work on superconductivity and magnetism at ultra-low
temperatures by Pobel's group in Germany \cite
{Pobell3,Pobell2,Pobell1,Pobell4}, as well as of Lounasmaa's one in Finland 
\cite{Loun1,Oja1}, at least two materials were discovered where S and
nuclear M order seem to coexist. The first one is metallic Rh, which is
superconducting at $T_S=325$ $\mu $K and whose nuclear moments might be
ordered antiferromagnetically at $T_N\sim 1$ $n$K, see Refs.~%
\onlinecite{Buchal,Hakonen}. There are also some hints on the AF order at
negative nuclear temperature $T_n$. Rh is an interesting system because of
its rather large Korringa constant $\kappa (\equiv \tau _1T_e)\leq 10$ s$%
\cdot $K, where $\tau _1$ is the spin-lattice relaxation time and $T_e$ is
the electronic temperature. Large $\kappa $ (or $\tau _1$) in Rh allows to
achieve very low nuclear temperatures $T_n\ll T_e$, as well as a realization
of negative $T_n$. The problem of the competition of nuclear magnetism and S
order in Rh will be studied elsewhere.

A remarkable achievement in this field was recently done by Pobel's group by
investigating AuIn$_2$, where the coexistence of the nuclear ferromagnetism
and superconductivity ($T_S=207$ $mK$) was found\cite{Pobell2},\cite{Pobell1}
below $T_M=35$ $\mu$K. Because of good thermal coupling of nuclear magnetic
moments to the conduction electrons in AuIn$_2$ (Korringa constant $\kappa
=0.1$ s$\cdot$K) the experiments were performed in thermal equilibrium $%
T_n=T_e$ down to $T=25$ $\mu$K. It was also found that AuIn$_2$ is a type-I
superconductor with the bare critical field $H_{c0}(T=0)=14.5$ G, which
would be in absence of the F ordering, while in its presence $H_c(T)$ is
decreased, i.e. $H_c=8.7$ G at $T=25$ $\mu$K. The latter result is a hint
that S and F orderings might coexist in the bulk down to $T=0$.

In the following the coexistence of S and M order in AuIn$_2$ is studied in
the framework of the microscopic theory of magnetic superconductors \cite
{BuBuKuPa1,BuBuKuPa2}. It considers interactions between LM's and conduction
electrons: a) via the direct hyperfine interaction -- because of simplicity
it is called the {\it exchange} (EX) one; b) via the dipolar magnetic field $%
{\bf B}_m({\bf r})=4\pi {\bf M}({\bf r})$, which is created by LM's, it is
called {\it electromagnetic} (EM) interaction. The general Hamiltonian which
describes conducting electrons and nuclei moments in AuIn$_2$ is given by 
\begin{eqnarray}
&&\hat H=\int d^3r\{\psi ^{\dagger }({\bf r})\epsilon ({\bf \hat p}-\frac{e}{c}{\bf %
A})\psi ({\bf r})+[\Delta ({\bf r})\psi ^{\dagger }({\bf r})i\sigma _y\psi
^{\dagger }({\bf r}) \nonumber \\
&&+{\rm c.c}]+\frac{|\Delta ({\bf r})|^2}V 
+\sum_iJ_{en}\delta ({\bf r}-{\bf r}_i)\psi ^{\dagger }({\bf r}){\bf %
\sigma I}_i\psi ({\bf r}) \nonumber \\ 
&&+\frac{[{\rm curl}{\bf A}({\bf r})]^2}{8\pi }\}  
+\sum_i[-{\bf B}({\bf r}_i)g_n\mu _n{\bf I}_i+\hat H_a({\bf I}_i)]+\hat H_{imp}.  
\label{eq.1}
\end{eqnarray}
Here, $\epsilon({\bf p})$, $\Delta ({\bf r})$, ${\bf A,}$
$J_{en}({\bf r})$ and $V$ are the quasiparticle energy, the superconducting
order parameter, the vector potential, the hyperfine contact coupling
between electronic spins ${\bf \sigma }$ (Pauli matrices) and localized
nuclear moments (LM's) ${\bf I}_i$ and the electron-phonon coupling constant
respectively. The first three terms in Eq.~(1) describe the superconducting
mean-field Hamiltonian in the magnetic field ${\bf B}({\bf r})={\rm curl}%
{\bf A}({\bf r})$ due to LM's and screening current, while the term $\hat H%
_{imp}$ describes the electron scattering (including also the spin-orbit
one) on nonmagnetic impurities. The term $-{\bf B}({\bf r}_i)g_n\mu _n{\bf I}%
_i$ describes the dipole-dipole interaction of LM's, as well as their
interaction with the magnetic field due to screening superconducting current
-- see more below. $\hat H_a({\bf I}_i)$ is (together with the dipole-dipole
interaction) responsible for magnetic anisotropy of LM's. In the case of AuIn%
$_2$, which is simple cubic crystal, its form is unknown -- see discussion
below. Later we show that under experimental conditions reported in Refs.~%
\onlinecite{Pobell3,Pobell2,Pobell1} the ferromagnetic structure, which
would be in absence of S order, is transformed in the presence of
superconductivity into spiral (or domain-like) one -- depending on magnetic
anisotropy.

A. The characteristic parameters of AuIn$_2$

The magnetic critical temperature $T_M$=35 $\mu$K is very small compared to $%
T_S\approx 0.2$ K, and it is much larger than the characteristic
dipole-dipole temperature $\Theta _{em}(\approx 1$ $\mu$K), see below. This
fact allows us to estimate the hyperfine contact interaction between
electrons and LM's, which is characterized by the parameter $%
h_{ex}=J_{en}(0)n_m\mid \langle{\bf I}_i\rangle\mid $, where $n_m$ is the
concentration of LM's. The indirect exchange energy (via conduction
electrons) between the LM's of nuclei is characterized by the RKKY
temperature $\Theta _{ex}=N(0)h_{ex}^2/n_m$, where $N(0)$ is the electronic
density of states at the Fermi level The crystallographic structure gives $%
n_m\approx 3\times 10^{22}$ cm$^{-3}$. $N(0)$ is obtained by knowing $%
H_{c0}(T=0)=14.5$ G, see Refs.~\onlinecite{Pobell3,Pobell2}, which gives $%
N(0)\approx 0.64\times 10^{34}$ erg$^{-1}$cm$^{-3}$. Since $T_M$(=35 $\mu$K)
is predominantly due to the indirect exchange interaction between In nuclei
moments one has $T_M\approx \Theta _{ex}$, which gives $h_{ex}\approx 1$ K.
Note that one has $h_{ex}>\Delta _0(\approx 0.36$ K), 
which gives rise to a gapless quasiparticle spectrum in S state below $T_M$
in clean samples ($l>\xi _0$) of AuIn$_2$, see below.

The electromagnetic (EM) dipole-dipole interaction between LM's is
characterized by $\Theta _{em}=2\pi n_m\mu ^2$, where $\mu =g_n\mu _nI.$ In
case of the In nuclei in the AuIn$_2$ cubic crystal one has $\mu \simeq
5.5\mu _n$, i.e. $\Theta _{em}\approx 1.2$ $\mu$K $(\ll T_M)$, which means
that the dipole-dipole interaction does not contribute to $T_M$ in AuIn$_2$.
However, it makes the magnetic structure transverse in S state, see below.

 From $\Delta _0(\simeq 1.76$ $T_S)$ and $v_F\approx 1.68\times 10^8$ cm/s
\cite{Pobell2,Pobell1} one gets $\xi _0\simeq 10^5$ $A$, while from the resistivity measurements 
\cite{Pobell3,Pobell1}, where RRR=500, one obtains $l\approx 3.6\times 10^4$ 
$A$. Accordingly, the spin-orbit scattering mean-free path is very large,
i.e. $l_{so}>3.6\times 10^4$ $A$, because one always has $l_{so}>l$. Note
that $l<\xi _0$ and the system is in the dirty (but not very dirty) limit.
The London penetration depth $\lambda _L\approx 200$ $A$\ \ is estimated
from $H_{co}$ and by knowing $\xi _0$ and $l$, which means that AuIn$_2$ is
the type I superconductor at temperatures where S and M orderings coexist.
>From the above analysis we estimate the parameter $(h_{ex}\tau /\hbar
)^2=0.1 $. It is small and dirty limit may be used to treat effect of
exchange field on superconductivity. This simplifies the theoretical
analysis given below. Here $\tau =l/v_F$ is the electron scattering time.

B. Theoretical analysis of AuIn$_2$

It was shown in Refs.~\onlinecite{BuBuKuPa1,BuBuKuPa2,Kauf} that when the
electron spin-orbit interaction is weak, i.e. when $l_{so}/l$ $\gg
(k_F^{-1}\xi _0/l^2)^{2/3}$, there is a peak in the spin susceptibility in
the superconducting state at nonzero wave vector $Q$. This means that in
the superconducting state an oscillatory magnetic order is more favorable
than the F one. In AuIn$_2$ one has $k_F=1.45$\AA$^{-1}$ and thus the condition for
an oscillatory magnetic order is $l_{so}$ $\gg 10^{-2}l$. Since 
by definition $l_{so}>l$ we see that the spin-orbit interaction does not
play any role in the formation of the magnetic structure in the coexistence
phase of AuIn$_2$. The magnetic order can be a spiral or domain-like one,
depending on the magnetic anisotropy, see below. The problem is now reduced
to the study of electrons moving in an oscillatory (with the wave vector $%
{\bf Q}$) exchange and magnetic field ${\bf h}_{ex}({\bf R})=h_{ex}{\bf S}(%
{\bf R})$, ${\bf B}({\bf R})={\rm curl}{\bf A}({\bf R})$ respectively. By
using the Eilenberger equations for the normal $g_\omega ({\bf v},{\bf R})$
and anomalous $f_\omega ({\bf v},{\bf R})$ electronic Green's function,
where the superconducting order parameter $\Delta ({\bf R})$ is a solution
of the self-consistency equation $\Delta ({\bf R})=2\pi g\sum_\omega \int d%
{\bf n}f_\omega ({\bf R},{\bf n})/4\pi $ ($g=N(0)V$ is the electron-phonon
coupling constant) one obtains the free-energy functional of the system $%
F_{SM}\{\Delta ,{\bf S}_Q,{\bf Q}\}$. It may be presented as a sum of
magnetic ($F_M$), superconducting ($F_S$) and interacting ($F_{int}$) parts,
i.e. $F_{SM}\{\Delta ,{\bf S}({\bf R})\}=F_S\{\Delta \}+F_M\{{\bf S}_Q,{\bf Q%
}\}+F_{int}\{\Delta ,{\bf S}_Q,{\bf Q}\}$). By assuming that: a) $\pi
E_S/F_{int}>1$, $E_S=N(0)\Delta ^2/2n_m$ -- this is indeed fulfilled in AuIn$%
_2$, where $\pi E_S/F_{int}\sim 100$, and b) the Fermi surface is isotropic
-- it is also fulfilled in AuIn$_2$, one gets the free-energy $F_{SM}$ 
\begin{eqnarray}
&&F_S\{\Delta \}=-\frac 12N(0)\Delta ^2\ln \frac{%
e\Delta _0^2}{\Delta ^2},  \\
&&F_{int}=F_{int}^{ex}+F_{int}^{em}=F_{int}^{ex}+\sum_{{\bf Q}}\frac{3\pi
^2\Theta_{em}\Delta \mid {\bf S}_{Q,_{\perp }}\mid ^2}{v_FQ(\lambda _LQ)^2}. \nonumber
\end{eqnarray}
Here the terms $F_{int}^{ex}$
and $F_{int}^{em}$ in Eq.~(2) describe the exchange EX and EM interaction of
superconducting electrons with LM's respectively. $F_{int}^{ex}$, and $F_M$
depend on $h_{ex}$, $S_Q$, $l$, $\xi _0$ etc. We consider only those cases
which might be important for the physics of AuIn$_2$.

$1.$ \underline{Dirty case ($l<\xi _0$)}. This case is {\it realized} in
AuIn$_2$ as reported in Refs.~\onlinecite{Pobell3,Pobell2}, where $l\approx
3.6\times 10^4$ $A$\ \ and $\xi _0\approx 10^5$ $A$. Immediately below the
magnetic critical temperature $T_M=35$ $\mu $K the magnetization is small
and $F_M$ has the form 
\begin{eqnarray}
&&F_M=\sum_{{\bf Q}}\{-\Theta _{ex}\chi _m^{-1}(Q)[\mid {\bf S}_{Q,\perp }\mid
^2+\mid {\bf S}_{Q,\parallel }\mid ^2] \nonumber \\
&&+\Theta _{em}\mid {\bf S}_{Q,\parallel
}\mid ^2\}+F_0+F_a
\end{eqnarray}
where ${\bf Q}\cdot {\bf S}_Q=QS_{Q,\parallel }$ and $\chi
_m^{-1}(T,Q)=(T-\Theta _{ex})/\Theta _{ex}+Q^2/12k_F^2$. Here, $k_F=1.45$ $%
A^{-1}$ is the Fermi wave vector. The isotropic term $F_0\{{\bf S}_Q^2\}$
(per LM) describes higher order terms in ${\bf S}_Q^2$, while $F_a$ (per LM)
describes magnetic anisotropy of the system, see discussion below.

Since the sample of AuIn$_2$ is in dirty limit\cite{Pobell3} -- it holds also 
$Ql\gg 1$, see below, $(h_{ex}\tau /\hbar )^2\approx 0.1\ll 1$, the
mean-free path drops out from the term $F_{int}^{ex}$ in Eq.~(2). Since we
consider the spiral (helical) structure with the amplitude $S_Q$, which
contains only one harmonic $Q$. Then the sum over ${\bf Q}$ in Eqs.~(2),(3)
drops out and one obtains $F_{int}^{ex}=\pi \Delta \cdot S_Q^2\Theta
_{ex}/2v_FQ$. By minimizing $F_{SM}\{\Delta ,S_Q,Q\}$ with respect to $%
\Delta ,S_Q$ and $Q$ one gets the {\it spiral} magnetic structure at $T$
very near $T_M$ with the wave vector $Q_S=(3\pi k_F^2/\xi _0)$ $^{1/3}$ $%
\approx 5\times 10^{-2}$ $A^{-1}$ and the period is $L_S=2\pi /Q_S\approx
120 $ $A$ - see $Fig.1a$. Note, one has $Q_Sl\sim 10^3$,
 i.e. the required theoretical
condition $Ql\gg 1$ is fulfilled in AuIn$_2$. In this case the EM
interaction in Eq.~(2) is negligible, i.e. $F_{int}^{em}/F_{int}^{ex}\approx
(\Theta _{em}/\Theta _{ex}\lambda _L^2Q^2)<10^{-2}$ and the spiral magnetic
structure is due to the effective EX interaction between electrons and LM's
(of In nuclei) mainly. However, the magnetic dipole-dipole interaction
between In nuclei, although small ($\Theta _{em}\ll \Theta _{ex}$), makes
the structure transverse (${\bf Q\cdot S}_Q=0$) due to the term $\Theta
_{em}\mid {\bf S}_{Q,\parallel }\mid ^2$ in Eq.~(3).

On cooling, $S_Q^2$ grows and higher order terms of $S_Q^2$ must be included
in $F_M$ as well as magnetic anisotropy $F_a$. However, in AuIn$_2$, which
is simple cubic structure, only higher order terms contribute to $F_a$ (per
LM), i.e. $F_a=D(S_x^4+S_y^4+S_z^4)$. One expects that $D<\Theta _{em}$ (or
more realistic $D\ll \Theta _{em}$, see below). However, if $D$ fulfills
the condition $(D/\Theta _{ex})^{3/4}>0.25(k_F\xi _0)^{-1/2}$ (one should
have $D/\Theta _{ex}>10^{-3}$ in the case of AuIn$_2$) the spiral structure
is transformed into a striped one-dimensional transverse domain structure 
\cite{BuBuKuPa1,BuBuKuPa2} - see $Fig.1b$. The condition on $D$ means that the domain-wall
thickness should be much smaller than the period of the domain structure.
The magnetic energy (per LM's) in the case of the domain magnetic structure
is given by $F_M=F_0(S_Q^2)-\Theta _{ex}S_Q^2+\eta (S_Q^2,T)Q/\pi $, where $%
F_0(S_Q^2)$ contains terms of higher order in $S_Q^2$. Here $\eta
=k_F^{-1}\Theta _wS_Q^2$ is the surface energy of the domain wall, and $%
\Theta _w\approx 0.58(\Theta _{ex}D)^{1/2}$ in the case of rotating Bloch
wall (where $D\ll \Theta _{ex}$). The wave vector of the domain structure $%
Q_D$, obtained by minimizing $F_{SM}$, at $T\ll T_M$ is given by $Q_{D\text{ 
}}\approx 2(\Theta _{ex}/\Theta _wk_F\xi _0)^{1/2}$. Note, the wave vector
of the domain structure $Q_{D\text{ }}$ is smaller than $Q_S$ for the
spiral. Because of the simple cubic structure of AuIn$_2$ it is more
probable that $D/\Theta _{ex}<10^{-3}$ than opposite, which means that AuIn$%
_2$ is the strong candidate to be the first system where superconductivity
coexists with {\it spiral} magnetic order down to $T=0$, see below. Namely,
it turns out that at $T=0$ the magnetic energy per LM is $F_M\approx -\Theta
_{ex}=-35$ $\mu $K, while the superconducting energy per LM is $F_S\approx
-4 $ $\mu $K. However, the interaction energy (per LM) is very small, $%
F_{int}\ll 10^{-2}$ $\mu $K. This result means that even at $T=0$ the loss
of energy due to the interaction, $F_{int}$, is much smaller than the gain
due to the condensation energy $F_S$, i.e. $F_S+F_{int}\approx F_S<0$, and
the oscillatory magnetic structure and S order coexist in AuIn$_2$ down to $%
T=0$.\\[-0.8cm]
\begin{figure}[tbp]
\epsfysize=2.8in 
\hspace*{2.0cm}
\epsffile{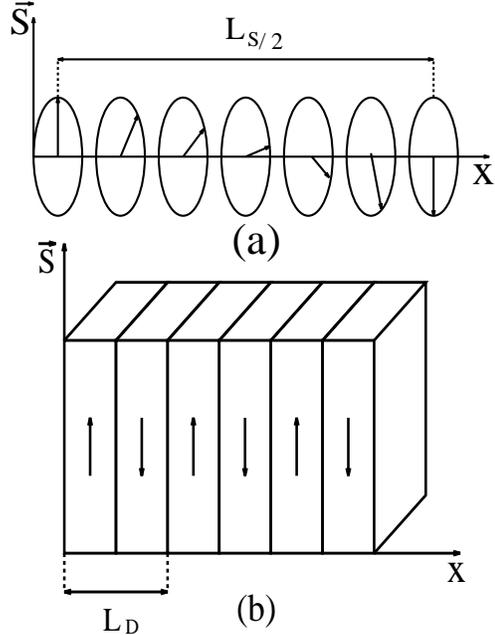}\\[-2.4cm]
\epsfysize=2.5in
\hspace*{1.6cm}
\epsffile{fig1b.ps}\\[-3cm]
\caption{
%Possible magnetic structures in the superconducting phase
%of $AuIn_2$: (a) spiral - for magnetic aniisotropy 
%$D/\Theta_{ex}<10^{-3}$; (b) domain - for $D/\Theta_{ex}>10^{-3}$.
(a) The spiral magnetic structure ${\bf S}(x)$, 
with the period $L_{S}$ in the 
superconducting phase for small anisotropy $D/\Theta _{ex}<10^{-3}$. 
(b) The domain-like magnetic structure ${\bf S}(x)$, 
with the period $L_{D}$ in the 
superconducting phase for appreciable anisotropy 
$D/\Theta _{ex}>10^{-3}$.
}
\end{figure}

$2$. \underline{Clean limit ($l>\xi _0$) }. The present experiments were
performed on dirty (but not very dirty) AuIn$_2$ samples, where $l<\xi _0$.
In that case the motion of Cooper pairs in the coexistence phase is
diffusive and there is an isotropization of the quasiparticle spectrum. This
means that the oscillatory magnetic structure acts like magnetic impurities
-- for similarity and differences of effects of magnetic impurities and the
oscillatory magnetic structure see Ref.~\onlinecite{BuBuKuPa1}. However, it
would be interesting to perform experiments on clean AuIn$_2$ samples with $%
l>\xi _0$ -- for instance on samples with a residual resistivity ratio RRR$>1500$. 
Namely, it was shown in Refs.~\onlinecite{BuBuKuPa1,BuRuKu} that the
oscillatory magnetic order in clean superconductors can give rise to the
gapless quasiparticle spectrum with nodes on a line at the Fermi surface if $%
h_{ex}>\Delta _0$. This is just the case in AuIn$_2$, where $h_{ex}\approx 1$
K and $\Delta _0\approx 0.36$ K. In the clean limit the quasiparticle motion
is anisotropic in the presence of an oscillatory magnetic structure with the
wave vector ${\bf Q}$, and the quasiparticle energy vanishes on lines at the
Fermi surface given by ${\bf v}_F\cdot {\bf Q}=0$ if $h(T)=h_{ex}S_Q(T)>%
\Delta _0$, see Ref.~\onlinecite{BuBuKuPa1}. In this case the density of
states for $E<\Delta _0$ is $N_s(E)=N(0)(\pi Eh/\Delta _0v_FQ_d)\ln (4\Delta
_0/\pi E)$ for the domain structure and $N_s(E)=N(0)(\pi Eh/\Delta _0v_FQ_d)$
for the spiral one. $N_s(E)$ can be experimentally obtained by measuring
voltage dependence of the tunneling conductivity in the S--N junction with
AuIn$_2$ being in the coexistence phase.

C. Effect of magnetic field

Because of very small interaction energy one expects that the critical field 
$H_c(T)$ does not vanish down to $T=0$. Indeed, equating Gibbs energy for 
superconducting state and that of normal ferromagnetic state one gets 
\begin{equation}  \label{eq.6}
F_S+F_M+F_{int}=F_M^0-\frac{H_c^2}{8\pi }-{\bf M\cdot H}_c,
\end{equation}
and if one defines $H_{SM}\equiv [8\pi (F_M^0-F_M-F_S-F_{int})]^{1/2}$ 
\begin{equation}  \label{eq.7}
H_c=\sqrt{H_{SM}^2+(4\pi M_0)^2}-4\pi M.
\end{equation}
At $T\ll T_M$ the magnetization ${\bf M}$ is saturated, i.e. $M\approx
M_0=5.5n_m\mu _n$. Because $F_{int}\ll F_S,F_M$ one has small difference in 
magnetic energy of oscillatory state, $F_M$ and that of ferromagnetic state, $F_M^0$. 
As result, $H_{SM}\approx
H_{c0}(0)=[8\pi (-F_S^0)]^{1/2}$. The experimental values \cite
{Pobell3} are $H_{c0}(0)\approx 14.5$ G and $4\pi M_0\approx 11$ G which
gives $H_c(0)\approx 7$ G. It was found experimentally \cite{Pobell3} 
that $H_c\approx 8.7$ G at $T=25$ $\mu $K, and
thus our estimate is reasonable. Nonzero value of $H_c(0)$ in AuIn$_2$ is
in a contrast to the case of magnetic type II superconductors ErRh$_4$B$_4$
and HoMo$_6$S$_8$, where $H_c(T)$ tends to zero [as well as $H_{c2}(T)]$ at $%
T\rightarrow T_M$, because in these compounds one has $F_{int}\approx|F_N-F_S|$ 
near some temperature $T_{S2}<T_M$. At $T>T_M$ one obtains $%
H_c=H_c^0/(1+4\pi \chi _M)$ in absence of demagnetization effects, where $%
\chi _M(T)=\bar \Theta _{em}/(T-\Theta _{ex})$ and $\bar \Theta _{em}=\Theta
_{em}/6\pi .$ The change of
the slope of $H_c(T)$ takes place at $T$ very near $T_M$. The experimental
broadening of the transition in the magnetic field can be due to the
polycrystalinity of the sample, where even small magnetic anisotropy of the
crystallites can produce percolation-like broadened resistive transition in
magnetic field \cite{BuBuKu,BuBu}.

In conclusion, we found, that superconductivity coexists with a domain-like
magnetic structure if the anisotropy parameter $D$ is not too small, i.e. $%
D/\Theta _{ex}>10^{-3}$. We estimate the period $L_D\approx 300$ \AA \ \ for $D\sim
1 $ $\mu$K. In the opposite case, $D/\Theta _{ex}<10^{-3}$, the magnetic
structure is spiral with the period $L_S\approx 120$ \AA. The realization of
the spiral structure in AuIn$_2$ is more probable due to the simple cubic
structure of this compound and accordingly due to small magnetic anisotropy.

It is also proposed that in the case of very clean AuIn$_2$ samples with RRR$>1500$ 
there is a line, given by ${\bf v}_F\cdot {\bf Q}=0$, at the Fermi
surface with nodes in the quasiparticle spectrum in the coexistence phase.
It would be interesting to study this regime experimentally, because in that
case thermodynamic and transport properties show power law behavior.

We would like to devote this paper to the pioneer in the field of magnetic
superconductors Vitalii Lazarevich Ginzburg on the occasion of his
80-th aniversary.
M.L.K. acknowledges Universit\'e Bordeaux for kund hospitality and O. Andersen, L.
Hedin, H.-U. Habermeier, C. Irslinger, Y. Leroyer, M. Mehring, K.-D. Schotte
and V.S.~Oudovenko for support.\\[-0.5cm]


\begin{references}
\bibitem{Pobell3}  S. Rehmann, T. Hermannsd\"orfer and F. Pobel, Phys. Rev.
Lett. {\bf 78}, 1122 (1997).

\bibitem{Ginzburg}  V.L. Ginzburg, Zh. Eksp. Teor. Fiz. {\bf 31}, 202 (1956).

\bibitem{BuBuKuPa1}  L.N. Bulaevskii, A.I. Buzdin, M.L. Kuli\'c and S.V.
Panyukov, Adv. Phys., {\bf 34}, 175 (1985); Sov. Phys. Uspekhi {\bf 27}, 927
(1984).

\bibitem{Maple}  M.B. Maple, H.C. Hammaker and L.D. Woolf, {\it in
Superconductivity in Ternary Compounds II, Topics in Current Physics, ed.
M.B. Maple and \O. Fischer, Springer Verlag}, v.~34, (1982).

\bibitem{Buzdin}  A.I. Buzdin, L.N. Bulaevskii, Sov. Phys. Uspekhi {\bf 29},
412, (1986).

\bibitem{BuBuKuPa2}  L.N. Bulaevskii, A.I. Buzdin, M.L. Kuli\'c and S.V.
Panyukov, Phys. Rev. B {\bf 28}, 1370 (1983).

\bibitem{BuRuKu}  L.N. Bulaevskii, M.L. Kuli\'c and A.I. Rusinov, Solid
State Comm., {\bf 30}, 59 (1979); J. Low Temp. Phys., {\bf 39}, 256 (1980).

\bibitem{BloVar}  E.I. Blount and C.M. Varma, Phys. Rev. Lett. {\bf 42},
1079 (1979).

\bibitem{Chang}  L.J. Chang, C.V. Tomy, D.M. Paul and C. Ritter, Phys. Rev.
B {\bf 54}, 9031 (1996).

\bibitem{Pobell2}  T. Hermannsd\"orfer and F. Pobel, J. Low Temp. Phys., 
{\bf 100}, 253 (1995).

\bibitem{Pobell1}  T. Hermannsd\"orfer, P. Smeibidl, B. Schr\"oder-Smeibidl
and F. Pobel, Phys. Rev. Lett. {\bf 74}, 1665 (1995).

\bibitem{Pobell4}  F. Pobel, Physics Today, January 1993; Physikalische
Bl\"atter, {\bf 50}, 853 (1994); Physica B {\bf 197}, 115 (1994).

\bibitem{Loun1}  O.V. Lounasmaa, Physics Today, October 1989; P. Hakonen and
O.V. Lounasmaa, Science {\bf 265}, September 1994; P. Hakonen, O.V.
Lounasmaa and A. Oja, J. Magn. and Magn. Mat. {\bf 100}, 394 (1991).

\bibitem{Oja1}  A.S. Oja and O.V. Lounasmaa, Rev. Mod. Phys., {\bf 69}, 1
(1997).

\bibitem{Buchal}  Ch. Buchal, F. Pobel, R.M. Mueller, M. Kubota and J.R.
Owers-Bradley, Phys. Rev. Lett. {\bf 50}, 64 (1983).

\bibitem{Hakonen}  P.J. Hakonen, R.T. Vuorinen, and J.E. Martikainen, Phys.
Rev. Lett. {\bf 70}, 2818 (1993).


\bibitem{Kauf}  M. Kaufman and O. Entin-Wohlman, Physica, B{\bf 69}, 77
(1976).

\bibitem{BuBuKu}  L.N. Bulaevskii, A.I. Buzdin, M.L. Kuli\'c, Solid State
Com., {\bf 41}, 309 (1981); Phys. Lett. A {\bf 85}, 169 (1981).

\bibitem{BuBu}  A.I. Buzdin and L.N. Bulaevskii, Fiz. Nizkih, Temp. {\bf 6},
1528 (1980).
\end{references}
\end{document}